\documentclass[journal=jacsat,manuscript=article]{achemso}

\usepackage{bm}
\usepackage{amssymb}
\usepackage{graphicx}
\usepackage{amsmath}
\usepackage{textcomp}
\usepackage{colordvi}
\usepackage{multirow}
\usepackage{ulem}
\usepackage{tikz}
\usepackage{color}
\usepackage{makecell}
\usepackage{booktabs}
\usepackage{braket}

\SectionNumbersOn

\usepackage{chemformula} 
\usepackage[T1]{fontenc} 

\author{Kainan Chang}
\email{knchang@ciomp.ac.cn}
\affiliation{Key Laboratory of Luminescence Science and Technology , Chinese Academy of Sciences \& State Key Laboratory of Luminescence Science and Applications, Changchun Institute of Optics, Fine Mechanics and Physics, Chinese Academy of Sciences.
}
\alsoaffiliation{University of Chinese Academy of Science, Beijing 100039, China.}

\author{Muhammad Zubair}
\affiliation{Department of Physics and Astronomy, The University of Alabama, Alabama 35487, USA.}

\author{Jin Luo Cheng}
\email{jlcheng@ciomp.ac.cn}
\affiliation{Key Laboratory of Luminescence Science and Technology , Chinese Academy of Sciences \& State Key Laboratory of Luminescence Science and Applications, Changchun Institute of Optics, Fine Mechanics and Physics, Chinese Academy of Sciences.
}
\alsoaffiliation{University of Chinese Academy of Science, Beijing 100039, China.}

\author{Wang-Kong Tse}
\email{wktse@ua.edu}
\affiliation{Department of Physics and Astronomy, The University of Alabama, Alabama 35487, USA.}

\title{Second Harmonic Generation in Topological Insulators under Quantizing Magnetic Fields}

\abbreviations{IR,NMR,UV}

\begin{document}







\begin{abstract}
We theoretically investigate the second harmonic generation (SHG) of topological insulator surface states in a perpendicular magnetic field. Our theory is based on the microscopic expression of the second-order magneto-optical conductivity developed from the density matrix formalism, taking into account hexagonal warping effects on the surface states' band structure.
Using numerically exact Landau level energies and wavefunctions including hexagonal warping, we calculate the spectrum of SHG conductivities under normal incidence for different values of magnetic field and chemical potential. The imaginary parts of the SHG conductivities show prominent resonant peaks corresponding to one-photon and two-photon inter-Landau level transitions. Treating the hexagonal warping term perturbatively, these transitions are clarified analytically within a perturbation theory from which approximate selection rules for the allowable optical transitions for SHG  are determined. 
Our results show extremely high SHG susceptibility that is easily tunable by magnetic field and doping level for topological surface states in the far-infrared regime, exceeding that of many conventional nonlinear materials.
This work highlights the key role of hexagonal warping effects in generating second-order optical responses and provides new insights on the nonlinear magneto-optical properties of the topological insulators.

\end{abstract}


\section{Introduction}
Topological insulators (TIs), a novel phase of matter, have garnered considerable attention in recent years owing to their unusual properties of simultaneously exhibiting insulating bulk states and conducting surface states \cite{TI_RMP1,Sinha2023,Rakhmanova2020,Jiang2023,Checkelsky2011,Yar2022}.
Specifically, three-dimensional (3D) TIs, exemplified by bismuth telluride (Bi$_2$Te$_3$) and bismuth 
selenide (Bi$_2$Se$_3$), exhibit spin-momentum-locked surface states and strong spin-orbit coupling \cite{Wang2015,Fan2014,Yazyev2010}, 
and become potential material candidates in optoelectronics,  spintronics and quantum computing \cite{Garate2010,Hasan2010,Fu2008}.
A noteworthy aspect of the surface states in 3D TIs is their unique interaction with electromagnetic fields in second harmonic generation (SHG) \cite{McIver2012,Li2019,Sinha2023,Glinka2015}, high-order harmonic generation (HHG) \cite{Jiang2023,Bai2021}, surface photocurrent \cite{Braun2016,Yao2023}, and surface photogalvanic effects \cite{Yu2020,Wu2022}.

The existence of SHG requires the breaking of inversion symmetry. This can naturally happen at the surface termination of bulk inversion-symmetric materials, of which the bismuth chalcogenide class of TIs is an example. Two SHG contributions are known to exist \cite{Hsieh_PRB2011,McIver2012,Sinha2023,Wu2024}: one is due to the bulk electrons confined within a layer of accumulation region caused by band bending; the other is due to the topological surface state electrons. These two contributions can be experimentally distinguished and separately analyzed \cite{McIver2012,Sinha2023,Wu2024}. 
This work is solely focused on the SHG contribution due to the surface Dirac electrons, which is a signature of the nontrivial bulk band topology.

The low-energy behavior of the TI topological surface states near the Dirac point is commonly described by the massless Dirac Hamiltonian \cite{Zhang2009a,Liu_PRB2010}. However, this model is unable to theoretically capture SHG processes for normal light incidence because the Dirac cone energy dispersion has full rotational symmetry. 
The [111] surface of the Bi$_2$Se$_3$ and Bi$_2$Te$_3$ topological insulators have C$_{3\mathrm{v}}$ symmetry. At higher energies, the dispersion exhibits hexagonal warping  \cite{Fu2009}, showing a snowflake-like energy contour. The crucial role of band warping effects on SHG was recognized in a few early papers for two-dimensional hexagonal-lattice materials \cite{margulis2013,SHG_Gr}. 
Although this warping breaks the continuous rotational symmetry of the TI surface state Hamiltonian, the mirror symmetry 
with respect to the plane perpendicular to 
the $\Gamma K$ direction ($x$-direction) is still preserved, yielding a vanishing SHG along the $x$-direction. Therefore, the SHG contribution due to the topological surface states for frequencies within the bulk energy gap should occur only in the $y$-direction.

To induce a finite SHG response along both $x$- and $y$-directions, one can devise a strategy to break the inversion symmetry along both directions.  This could happen, for example, by illuminating the sample with circular polarized light \cite{McIver2012}, or by doping the 
system with magnetic impurities \cite{Li2019} where the combined effects of the magnetically-induced Dirac gap and the hexagonal warping give rise to an inversion-asymmetric energy dispersion along the $x$-direction. In this work, we propose a different strategy to induce SHG, by applying an external magnetic field.

Since the discovery of topological insulators, their linear magneto-optical responses have been intensively explored (see \textit{e.g.},   Refs.~\cite{Li_PRB2013,Wilson_PRB2014,Scharf_PRB2015}). One of the most dramatic effects is the quantized values of the Faraday and Kerr effects 
from the surface states of 3D TIs \cite{Tse2016,Tse2016a,Tse2010,Aguilar_PRL2012,wu2016quantized,okada2016terahertz,dziom2017observation}. Linear magneto-optical properties also provide valuable information on the responses of many emerging quantum materials under an external magnetic field  \cite{Tabert_PRB2013,Ashby_PRB2013,Tahir_PRB2015,Hien_PRB2020}. In contrast to linear responses, to our best knowledge there have been no theoretical or experimental studies on the nonlinear optical responses of 3D TIs in the presence of an external magnetic field.

This work aims to address this knowledge gap in our understanding of magneto-optical second harmonic generation. We perform a systematic study on the SHG from the topological surface states of 3D TIs in a perpendicular quantizing magnetic field. Our analysis is based on a fully microscopic expression of the nonlinear optical conductivity obtained within the density matrix formalism \cite{Cheng2018}. 
Using a perturbation scheme, the selection rules are first analytically identified for dipole matrix elements between different Landau levels, and then for the one- and two-photon resonant transitions contributing to SHG. Informed by these selection rules, we then analyze and interpret our numerical results of the SHG conductivity spectrum for different magnetic fields and chemical potentials.

This paper is organized as follows: In Sec.\,\ref{model}, we introduce the model for the TI surface states in a magnetic field and the second-order optical conductivity; in  Sec.\,\ref{results}, we first present a perturbation theory to obtain the electronic eigenstates and the selection rules, and then we present and discuss the results for the SHG spectra at different values of magnetic field and chemical potential; finally, Sec.\,\ref{conclusion} concludes our paper.

\section{Model}\label{model}

The surface states in Bi$_2$Se$_3$ can be determined by a Hamiltonian \cite{Fu2009,Li2019,Yar2022}
\begin{align}
H_{0}(\bm{p})=v_F(p_y\sigma_x-p_x\sigma_y)
+\frac{\lambda}{2\hbar^3}(p^3_+ + p^3_-)\sigma_z\,,
\label{H0}
\end{align}
where the first term at the right hand describes an isotropic two-dimensional massless Dirac Fermion with a Fermi velocity $v_F$, the second term describes the hexagonal warping effects with a strength parameter $\lambda$, $\sigma_{x,y,z}$ are Pauli matrices, and $\bm p$ is electron momentum with $p_{\pm}=p_x\pm ip_y$.
Applying a perpendicular magnetic field $\bm B=B\hat{\bm z}$, the Hamiltonian becomes
$H=H_0({\bm p+|e|\bm A})+\Delta\sigma_z$, where the vector potential $\bm A =(0,Bx,0)$ is chosen in the Landau gauge and $\Delta=g\mu_BB/2$ gives the Zeeman term with the g-factor $g$ and Bohr magneton $\mu_B$.
The Landau gauge maintains the translational invariance along the $y$-direction, and thus the energy eigenstates can be written as $\Psi(\bm r)={1}/{\sqrt{2\pi}}\exp(ik_y y)\Phi(x+l_c^2k_y)$, where $l_c=\sqrt{{\hbar}/{(eB)}}$ is the magnetic length, and  $\Phi(x)$ is a spinor envelope function satisfying
\begin{subequations}
\begin{align}
&\left[h^{(0)}+h^{(1)}\right]\Phi(x)=\varepsilon \Phi(x)\,,\label{hmatrix1}\\
&h^{(0)}=\frac{\hbar v_F}{\sqrt{2}l_c}
\left[
\hat{a}^\dagger(-i\sigma_x-\sigma_y)
+\hat{a}(i\sigma_x-\sigma_y)
\right]
+\Delta\sigma_z \,, \label{hmatrix2}
\\
&h^{(1)}=\sqrt{2}\lambda
\left(
\frac{\hbar}{l_c}
\right)^3
\left[
(\hat{a}^\dagger)^3+\hat{a}^3
\right]
\sigma_z \,. \label{hmatrix3}
\end{align}
\end{subequations}
Here $\hat{a}^\dagger=\frac{l_c}{\sqrt{2}\hbar}(p_x+ieBx)$ and $\hat{a}=\frac{l_c}{\sqrt{2}\hbar}(p_x-ieBx)$ are 
the raising and lowering operators with satisfying the commutation $[\hat{a},\hat{a}^\dagger]=1$. 
The solution of Eq.\,(\ref{hmatrix1}) gives the discrete Landau levels $\Phi_s(x)$ and eigenenergies $\varepsilon_s$ with the discrete quantum number $s$. 
The  $B$-dependence of eigenenergies are shown in Fig.\,\ref{energy}\,(a).
Without the warping term, $\Phi_s(x)$ can be analytically obtained \cite{Cheng2018}, and the quantum number can be taken as $s=\cdots,-2,-1,0,1,2,\cdots$; with the inclusion of the warping term, $\Phi_s(x)$ can be numerically solved by expanding it as $\Phi_s(x)=\sum_{n=0}^\infty\sum_{\sigma=\uparrow,\downarrow}C_{n\sigma}^s\phi_n(x)\chi_\sigma$, where $\chi_\sigma$ gives the spin eigenstate of $\sigma_z$, $\phi_n(x)$ with $n=0,1,2,\cdots$ is the $n$th harmonic oscillator eigenstate associated with the operator $\hat{a}$ and satisfies $\hat{a}\phi_0(x)=0$ and $\phi_n(x)=\hat{a}^\dagger \phi_{n-1}(x)/\sqrt{n}$ for $n>0$. 
With these eigenstates, the Berry connection $\bm\xi_{s_1s_2}$ between different Landau levels $\Phi_{s_1}(x)$ and $\Phi_{s_2}(x)$ can be calculated as $\bm\xi_{s_1s_2}=\sum_{\alpha=\pm}\xi^\alpha_{s_1s_2}\hat{\bm e}^{\bar\alpha}$ with
\begin{align}
\xi^+_{s_1s_2}=-il_c\int dx \Phi_{s_1}^\dag(x) \hat a^\dagger \Phi_{s_2}(x) = \left[\xi^-_{s_2s_1}\right]^\ast\,,
\label{berry}
\end{align}
where $\hat{\bm e}^\pm=(\hat{\bm e}^x\pm i\hat{\bm e}^y)/\sqrt{2}$ gives the circularly polarization vectors with the Greek index $\bar\alpha=+,-$ for $\alpha=-,+$.

We are interested in the SHG responding to a uniform electric field $\bm E(t)=\bm E_\omega e^{-i\omega t}+c.c.$ at a frequency $\omega$, and the associated response current is written as $\bm J_{\text{SHG}}(t)=\bm J_{2\omega}e^{-2i\omega t}+c.c.$. Usually the response coefficients $\sigma^{abc}(\omega)$ for $a,b,c=x,y$ are defined in Cartersian coordinates as $J_{2\omega}^a=\sum_{bc=x,y}\sigma^{abc}(\omega)E_\omega^bE_\omega^c$. 
Our focus in this work is on the response to a circularly polarized light. 
For the electric field $\bm E_\omega$ and the current density $\bm J_{2\omega}$, their circularly polarized components are defined as $E^\alpha_\omega=\bm E_\omega\cdot\hat{\bm e}^{\bar\alpha}$ and $J^\alpha_{2\omega}=\bm J_{2\omega}\cdot\hat{\bm e}^{\bar\alpha}$; note that these two definitions are different from that of $\xi^\alpha$. Then the SH current becomes $J^\tau_{2\omega}=\sum_{\alpha\beta=\pm}\sigma^{\tau\alpha\beta}(\omega)E_\omega^\alpha E_\omega^\beta$, where the microscopic response coefficients $\sigma^{\tau\alpha\beta}(\omega)$ is given as \cite{Cheng2018}
 \begin{align}
 \sigma^{\tau\alpha\beta}(\omega)
=&-i\frac{e^4B}{2\pi\hbar^2}
\sum_{s_1s_2s} 
\frac{\hbar\omega_{s_2s_1}\xi^{\bar\tau}_{s_2s_1}(\xi^{\alpha}_{s_1s}\xi^{\beta}_{ss_2}+\xi^{\beta}_{s_1s}\xi^{\alpha}_{ss_2})}
{2\hbar\omega-\hbar\omega_{s_1s_2}+i\Gamma} \notag\\
&\times
\left(\frac{f_{s_2s}}
{\hbar\omega-\hbar\omega_{ss_2}+i\Gamma}
-\frac{f_{ss_1}}
{\hbar\omega-\hbar\omega_{s_1s}+i\Gamma}\right)
\,,
\label{sigma}
\end{align}
with $\hbar\omega_{s_1s_2}=\varepsilon_{s_1}-\varepsilon_{s_2}$, $\Gamma$ phenomenological relaxation energy,  $f_s=[1+e^{(\varepsilon_s-\mu)/(k_BT)}]^{-1}$ the Fermi-Dirac distribution at chemical potential $\mu$ and temperature $T$, and $f_{s_1s_2}=f_{s_1}-f_{s_2}$. 

We first perform a symmetry analysis on the tensors of $\sigma^{\tau\alpha\beta}$ and $\sigma^{abc}$.
When there is no external magnetic field, the surface states of Bi$_2$Se$_3$ have C$_{3\mathrm{v}}$ symmetry \cite{Fu2009}; an external magnetic field along the $z$ direction breaks the mirror symmetry with respect to the $x=0$ plane, and the surface states reduce into C$_{3}$ symmetry. 
Therefore, at nonzero magnetic field, the nonzero in-plane components of $\sigma^{dab}$ are $\sigma^{xyy}=\sigma^{yyx}=\sigma^{yxy}=-\sigma^{xxx}$ and $\sigma^{yxx}=\sigma^{xxy}=\sigma^{xyx}=-\sigma^{yyy}$;
then the nonzero components of $\sigma^{\tau\alpha\beta}=\sum_{abc=x,y} \left(\hat{\bm e}^{\bar\tau}\cdot\hat{\bm e}^a\right)\left(\hat{\bm e}^\alpha\cdot\hat{\bm e}^b\right)\left(\hat{\bm e}^\beta\cdot\hat{\bm e}^c\right) \sigma^{abc}$ are 
\begin{align}
{\sigma}^{+--}&=\sqrt{2}\left(\sigma^{xxx}+i\sigma^{yyy}\right)\,,\quad
{\sigma}^{-++}=\sqrt{2}\left(\sigma^{xxx}-i\sigma^{yyy}\right)\,.
\label{relation}
\end{align}
Therefore,  a pure circularly polarized light can generate SH current with the opposite circularly polarization.
At $B=0$, the mirror symmetry leads to $\sigma^{xxx}=0$ and $\sigma^{+--}=-\sigma^{-++}$.

\section{Results}
\label{results}

\subsection{Selection rules and Kramers-Kr\"onig relation}

\begin{figure*}
		\centering
  \includegraphics[scale=0.95]{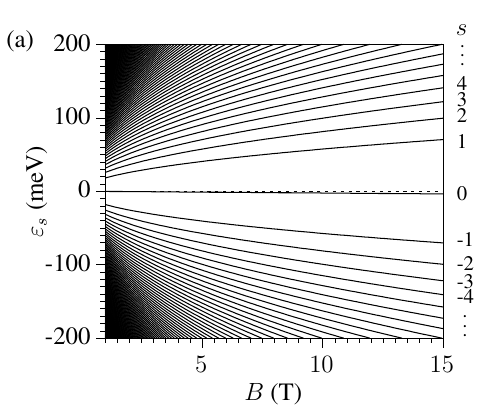}
  \includegraphics[scale=0.95]{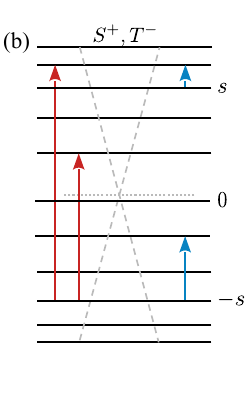}
  \includegraphics[scale=0.95]{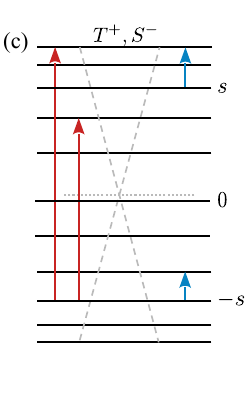}
\caption{
(a) Eigenenergies $\varepsilon_s$ of the Landau levels as a function of the magnetic field $B$. 
The index $s$ is also labeled. 
Selection rules for one- and two-photon resonant transitions between different Landau levels indicating by horizontal lines 
(b) $S^+_{s_1s_2}$ and $T^-_{s_1s_2}$, and (c) $T^+_{s_1s_2}$ and $S^-_{s_1s_2}$.
The band structure at zero magnetic field is also illustrated.
The red (blue) arrows denote the interband (intraband) transitions.
}
\label{energy}
\end{figure*}

With the inclusion of the warping term, the Hamiltonian $h=h^{(0)}+h^{(1)}$ cannot be analytically diagonalized. 
When the warping parameter $\lambda$ in $h^{(1)}$ is small,  the electronic eigenstates and the selection rules for the Berry connection of $\xi^+$ can be obtained using a perturbation method. 
To better show the effect of the warping term, the $\lambda$ dependence of all quantities is explicitly shown as $\Phi_s(x,\lambda)$, $\varepsilon_s(\lambda)$, $\xi^+(\lambda)$, and so on. 
Without the warping term, the eigenstates are
\begin{align}
\Phi_s(x,0)=\frac{1}{\sqrt{2}}\left[ig_s\sqrt{1+{\cal N}_s}\phi_{|s|-1}(x)\chi_\uparrow+\sqrt{1-{\cal N}_s}\phi_{|s|}(x)\chi_\downarrow\right]\,,
\label{eq-state}
\end{align} 
and the eigenenergies are $\varepsilon_s(0)=g_s \sqrt{\Delta^2+|s|(\hbar\omega_c)^2}$ with $g_s=\text{sgn}(s)$ for $s\neq0$ and $g_0=-1$, $\mathcal{N}_{s}=\Delta/\varepsilon_{s}(0)$, and the cyclotron energy $\hbar\omega_c=\sqrt{2}\hbar v_F/l_c=\sqrt{2\hbar eB}v_F$.
Further the optical dipole matrix elements are  $\xi_{s^\prime s}^+(0)=-\frac{il_c}{2}R_{s^\prime s}\delta_{|s^\prime|,|s|+1} $ with a selection rule $|s^\prime|=|s|+1$ and 
\begin{align}
R_{s^\prime s}= &g_{s^\prime}g_{s}\sqrt{(1+{\cal N}_{s^\prime})(1+{\cal N}_{s})}\sqrt{|s|}
+\sqrt{(1-{\cal N}_{s^\prime})(1-{\cal N}_{s})}\sqrt{|s|+1}\,.
\label{eq-app-xi0}
\end{align}
It is easy to check that $R_{s^\prime s}$ is always a positive number.

Up to the first order, the eigenstates can be written as
\begin{align}
    \Phi_s(x,\lambda) \approx \Phi_s(x,0) 
    + \lambda \sum_{s^\prime,\pm} \Phi_{s^\prime}(x,0) A_{s^\prime s}^\pm \delta_{|s|\pm3,|s^\prime|}\,,
\end{align}
with the first order correction coefficient
\begin{align}
A_{s^\prime s}^\pm
=\frac{
\sqrt{2}\hbar^3
\left\{
\pm[\xi^{\pm}(0)]^3_{s^\prime s}{\cal N}_s
\mp[\xi^{\pm}(0)]^3_{s^\prime -s}\sqrt{1-{\cal N}_s^2}
\right\}
}
{il_{c}^{4}[\varepsilon_{s}(0)-\varepsilon_{s^\prime}(0)]}
\,,
\end{align}
which is always a real number.
It is easy to check that the first-order energy correction is zero, thus
\begin{align}
\varepsilon_s(\lambda)=\varepsilon_s(0)+O(\lambda^2)\,.
\end{align}
In this case, the Berry connection $\xi_{s_1s_2}^+(\lambda)$ is
\begin{align}
\xi^+_{s_1s_2}(\lambda)
 \approx 
{\xi}_{s_1s_2}^{+}(0) 
+ \lambda \sum_{s^\prime,\pm} 
\left[(A_{s^\prime s_1}^{\pm})^\ast 
\xi_{s^\prime s_2}^+(0) 
+
\xi_{s_1s^\prime}^+(0)A_{s^\prime s_2 }^{\pm}
\right]
\delta_{|s_1|\pm3,|s_2|+1}
\,.
\end{align}
Thus $\xi_{s_1s_2}^+(\lambda)$ is still a pure imaginary number. 
Therefore, using the selection rules for $\xi_{s_1s_2}^+(0)$, the first-order perturbative $\xi_{s_1s_2}^+(\lambda)$ is nonzero only between states satisfying the selection rule $|s_1|-|s_2|=3l+1$ with $l=-1,0,1$. 
Perturbatively, $l=0$ corresponds to the zero-order term of $\xi^+$, while $l=\pm 1$ corresponds to the first-order term of $\xi^+$.
Note that with the inclusion of all higher order perturbation terms the value of $l$ will be extended to all integers. 

With these selection rules, we can further analyze the SHG conductivity $\sigma^{\bar\tau\tau\tau}$. We first look at the expression in Eq.\,(\ref{sigma}) in the limit of $\Gamma\to0$, where the conductivity can be written as $\left.\sigma^{\bar\tau\tau\tau}(\omega)\right|_{\Gamma\to0}={\cal A}_r^\tau(\omega)+i{\cal A}_i^\tau(\omega)$ with
\begin{align}
    {\cal A}_i^\tau(\omega) &= \sum_{s_1s_2}\left[T_{s_1s_2}^{\tau}f_{s_2s_1}\delta(2\hbar\omega-\hbar\omega_{s_1s_2}) + S_{s_1s_2}^\tau f_{s_2s_1}\delta(\hbar\omega-\hbar\omega_{s_1s_2})\right]\,,\label{eq:s2}\\
    T_{s_1s_2}^\tau &= i\frac{2e^4B}{\hbar^2}\sum_s\frac{\hbar\omega_{s_2s_1}\xi^\tau_{s_2s_1}\xi^\tau_{s_1s}\xi^\tau_{ss_2}}{\varepsilon_{s_1}+\varepsilon_{s_2}-2\varepsilon_s}\,,\label{eq:T}\\
    S^\tau_{s_1s_2}&=i\frac{2e^4B}{\hbar^2}\sum_s\hbar\omega_{s_2s_1}\xi^\tau_{s_2s}\xi^\tau_{ss_1}\xi^\tau_{s_1s_2}\left(\frac{1}{2\varepsilon_{s_1}-\varepsilon_{s_2}-\varepsilon_s}+\frac{1}{2\varepsilon_{s_2}-\varepsilon_{s_1}-\varepsilon_s}\right)\,,\label{eq:S}\\
{\cal A}_r^\tau(\omega) &= \frac{1}{\pi}\int_{-\infty}^{+\infty}\frac{{\cal A}_i^\tau(\omega^\prime)}{\omega^\prime-\omega}d\omega^\prime
= \frac{1}{\pi}\int_{0}^{+\infty}\frac{{\cal A}_i^\tau(\omega+\omega^\prime)-{\cal A}_i^\tau(\omega-\omega^\prime)}{\omega^\prime}d\omega^\prime
\,.
\end{align}
Because $\xi_{s_1s_2}^\tau$ is a pure imaginary number, these quantities $S^\tau_{s_1s_2}$, $T^\tau_{s_1s_2}$, ${\cal A}_i^\tau(\omega)$, and ${\cal A}_r^\tau(\omega)$ are all real.
${\cal A}_{i}^\tau(\omega)$ is nonzero only at discrete photon energies. 
 The coefficients satisfy $S_{s_1s_2}^\tau =\left[S_{s_2s_1}^{\bar\tau}\right]^\ast$ and $T_{s_1s_2}^\tau =\left[T_{s_2s_1}^{\bar\tau}\right]^\ast$.
The coefficients $T^\tau_{s_1s_2}$ and $S^\tau_{s_1s_2}$ in Eq. (\ref{eq:s2}) denote the amplitudes for  two-photon and one-photon resonant processes, respectively, and they indicate
two types of resonances in the SHG conductivity: (1) The term involving $S^\tau_{s_1s_2}$ gives one-photon resonant transition from the Landau levels $s_2$ to $s_1$, occurring at photon energy $\hbar\omega=\hbar\omega_{s_1s_2}$. Using the selection rules of $\xi^\tau_{s_1,s_2}$, all involved electronic states in the one-photon resonant transition satisfy $|s_1|-|s_2|=3l_1+\tau$, $|s|-|s_1|=3l_2+\tau$, and $|s_2|-|s|=3l_3+\tau$ for integer $l_1$, $l_2$, and $l_3$ with a further condition $l_1+l_2+l_3+\tau=0$. Perturbatively, the first-order term of $S^\tau_{s_1s_2}$ requires that only one of $l_1$, $l_2$, $l_3$ to be $-\tau$, therefore the selection rules for one-photon resonant transition between the unoccupied level $s_1$ and the occupied level $s_2$ are $|s_1|-|s_2|=\tau$ or $-2\tau$. Up to the first-order term, all nonzero $S_{s_1s_2}^\tau$ are $S_{\pm(s+1),\pm s}^+$, $S_{\pm s,\pm(s+2)}^+$, $S_{\pm s,\pm(s+1)}^-$, and $S_{\pm (s+2),\pm s}^-$ for $s=0,1,2,\cdots$; among them $S_{s+1,\pm s}^+$, $S_{\pm s,-s-2}^+$, $S_{\pm s,-s-1}^-$, and $S_{s+2,\pm s}^-$ have positive transition energies. Taking the Landau levels $s\le0$ with energies $\varepsilon_s<0$ as valence band states and the Landau levels $s>0$ with energies $\varepsilon_s>0$ as conduction band states, these resonant transitions can be further divided into interband transitions $S^+_{s+1,-s}$, $S^+_{s+1,-s-3}$, $S^-_{s+1,-s-2}$, and $S^-_{s+2,-s}$, as well as the intraband transitions $S^+_{s+2,s+1}$, $S^+_{-s,-s-2}$, $S^-_{-s,-s-1}$, and $S^-_{s+3,s+1}$  for $s=0,1,2,\cdots$. 
(2) The term involving $T_{s_1s_2}^\tau$ gives two-photon resonant transition from Landau levels $s_2$ to $s_1$, occurring at photon energies $\hbar\omega=\hbar\omega_{s_1s_2}/2$. Similarly, up to the first-order perturbation the selection rule for this resonant transition between the unoccupied state $s_1$ and occupied state $s_2$ is $|s_1|-|s_2|=-\tau$ or $2\tau$. 
Up to the first order, the possible two-photon resonant transitions with positive transition energies are $T_{\pm s,-s-1}^+$, $T_{s+2,\pm s}^+$, $T_{s+1,\pm s}^-$, and $T_{\pm s,-s-2}^-$ for $s=0,1,2,\cdots$. The interband resonant transitions are $T^+_{s+1,-s-2}$, $T^+_{s+2,-s}$, $T^-_{s+1,-s}$, and $T^-_{s+1,-s-3}$; and the intraband resonant transitions are $T^+_{-s,-s-1}$, $T^+_{s+3,s+1}$, $T^-_{s+2,s+1}$, and $T^-_{-s,-s-2}$ for $s=0,1,2,\cdots$. 
These resonant transitions are illustrated in Figs.\,\ref{energy}\,(b) and (c).

%
%
By checking the expression in Eq.\,(\ref{sigma}), and extending the argument $\omega$ to the complex plane, all poles of the conductivity are in the lower half-plane, and $\sigma^{\bar{\tau}\tau\tau}(\omega)$ is an analytic function of $\omega$ in the upper half-plane. Because of $\sigma^{\bar\tau\tau\tau}(\omega\to\infty)\to0$, thus the general Kramers-Kr\"onig relation \cite{Boyd2008} is still satisfied
\begin{align}
	\text{Re}[\sigma^{\bar\tau\tau\tau}(\omega)] =\frac{1}{\pi} 
	\int^\infty_{-\infty} \frac{\text{Im}[\sigma^{\bar\tau\tau\tau}(\omega^\prime)]}{\omega^\prime-\omega} d\omega^\prime \,.
\end{align}
Therefore, the following discussions mainly focus on the imaginary part of the conductivity.

\subsection{SHG under magnetic field $B=5$ T}

\begin{figure*}
		\centering
        \includegraphics[scale=1]{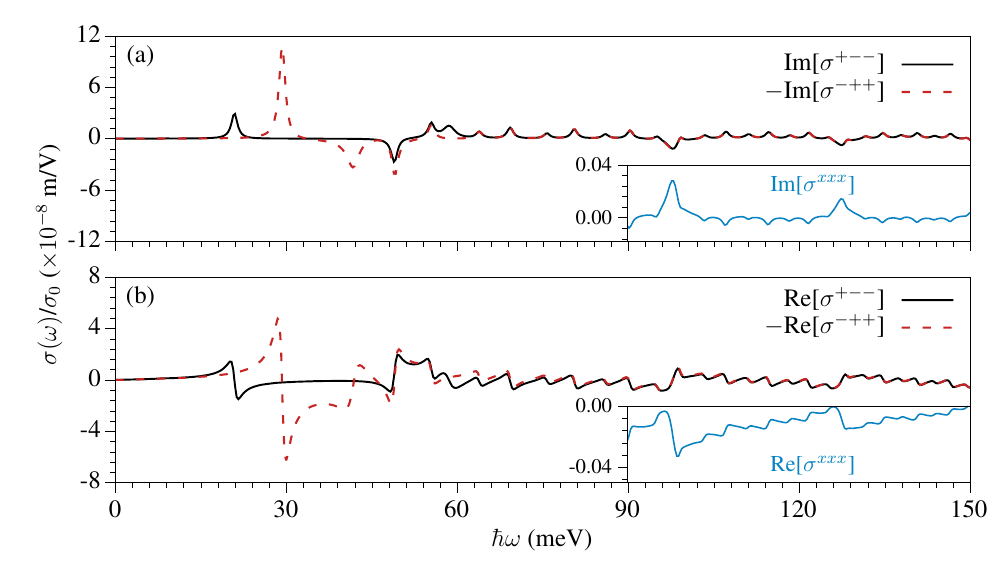}
         \includegraphics[scale=1]{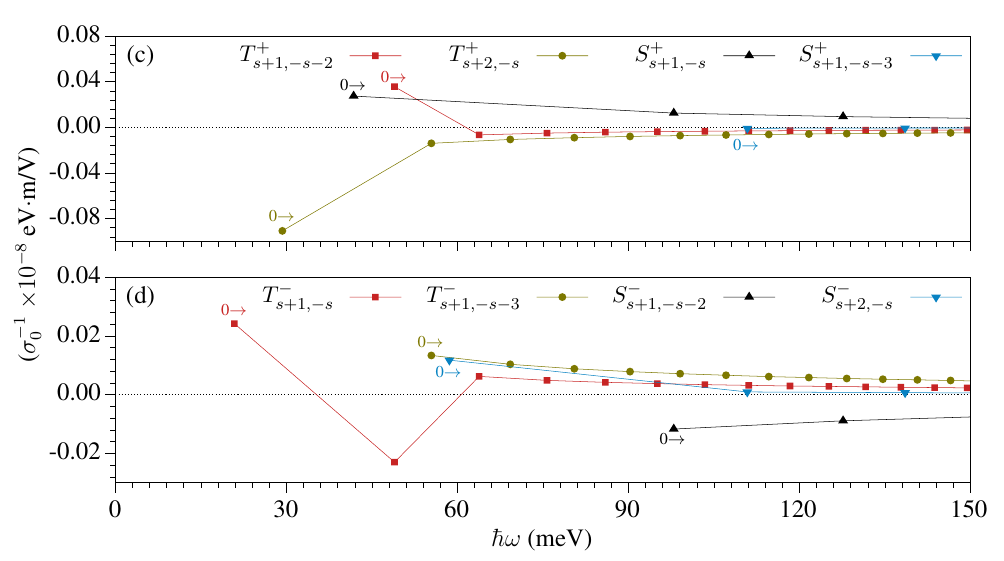}
\caption{
Spectra for SHG conductivity under magnetic field $B=5$ T: (a) imaginary part and (b) real part of  ${\sigma}^{+--}$ (black solid line) and $-{\sigma}^{-++}$ (red dashed line).
The inserts are spectra of 
${\sigma}^{xxx}=\frac{1}{2\sqrt{2}}({\sigma}^{+--}+{\sigma}^{-++})$.
Interband transition amplitudes (c) $T^-_{s+1,-s}$, $T^-_{s+1,-s-3}$, $S^-_{s+1,-s-2}$, $S^-_{s+2,-s}$,
(d) $T^+_{s+1,-s-2}$, $T^+_{s+2,-s}$, $S^+_{s+1,-s}$, and $S^+_{s+1,-s-3}$ 
for $s \ge 0$ are plotted as functions of the resonant transition energies. The quantities with $s=0$ are marked.
}
\label{shg}
\end{figure*}

In the current and the following sections, we obtain the exact SHG conductivities from Eq.~\eqref{sigma} by numerically diagonalizing the Hamiltonian Eqs.~\eqref{hmatrix1}-\eqref{hmatrix3} including the warping term. Throughout this work, we take the material parameters $v_F=5\times10^5$ m$\cdot$s$^{-1}$,  $\lambda=1.65\times10^{-28}$ eV$\cdot${m}$^3$, $\Gamma=1.3$ meV, $g=8.4$, and $T=1$ K. 

In this section, the magnetic field is taken as $B=5$ T, and the chemical potential is chosen so that the highest occupied state is $s=0$.
Figures\,\ref{shg}\,(a) and (b) show the spectra of  $\sigma^{+--}(\omega)$ and $\sigma^{-++}(\omega)$ in a photon energy range of $\hbar\omega\in[0,150]$ meV.
The spectra exhibit the following features: 
(1) The imaginary parts exhibit discrete peaks, which is consistent with our discussion in the previous section. Moreover, taking $\sigma^{-++}(\omega)$ as example, because the highest occupied state is $s=0$, all of these peaks are induced by resonant interband transitions $S^+_{s+1,-s}$, $S^+_{s+1,-s-3}$, $T^+_{s+1,-s-2}$, and $T^+_{s+2,-s}$ for $s=0,1,2,\cdots$, where the transition energies are $\hbar\omega=\varepsilon_{s+1}-\varepsilon_{-s}$, $\varepsilon_{s+1}-\varepsilon_{-s-3}$, $(\varepsilon_{s+1}-\varepsilon_{-s-2})/2$, and $(\varepsilon_{s+2}-\varepsilon_{-s})/2$, respectively. 
To better understand these transitions, interband $S^+$ and $T^+$ are plotted as functions of the resonant transition energies in Fig.\,\ref{shg}\,(c), where the minimal transition energy occurs at $s=0$, and the energy increases as $s$ increases.
It can be seen that for two-photon processes the first peak originates from the transition $T^+_{2,0}$ and the second peak from $T^+_{1,-2}$, and other two-photon peaks can also be identified.
Moreover, except $T^+_{1,-2}$ all other $T^+$ are negative, which are consistent with the peak directions in Fig.\,\ref{shg}\,(a). 
For one-photon processes, the peaks from $S^+_{s+1,-s}$ are obvious with positive amplitudes, while those from $S^+_{s+1,-s-3}$ have too small amplitudes to be seen in the figure.
At this chemical potential all of the peaks come from interband transitions.
Similarly, the transition amplitudes $S^-$ and $T^-$ for $\sigma^{+--}(\omega)$ are shown in Fig.\,\ref{shg}\,(d) and can be used to analyze the spectra of $\sigma^{+--}(\omega)$.
(2) With the increase of the photon energy, the energy difference between neighbor peaks decreases. The resonant peaks at high photon energies are induced by resonant optical transitions between high Landau levels, for which the energy levels become denser. 
(3) For high photon energies ($\hbar\omega\gtrsim60$~meV) the values of $\sigma^{+--}$ and $\sigma^{-++}$ are approximately opposite. This can be understood as following: The energies of Landau levels approximately satisfy  $|\varepsilon_{s}|\approx|\varepsilon_{-s}|$ for $s>0$, thus the transitions, taking $T_{s,-s-1}^+$ and $T_{s+1,-s}^-$ as example, have almost the same transition energies. When these transitions do not involve the $s=0$ state, their amplitudes are about opposite. However, they are still different, 
because their sum gives nonzero $\sigma^{xxx}=\frac{1}{2\sqrt{2}}(\sigma^{+--}+\sigma^{-++})$, which is shown in the inset.
Note, the first several peaks show significant differences because these peaks involve the Landau level $s=0$, which has no corresponding level with opposite energy.
(4) While the imaginary part of the spectra shows discrete peaks, the real part, which can be obtained using the Kramers-Kr\"onig relation, crosses zero  at the corresponding locations where the imaginary part peaks. 
These behaviors are opposite to what occur in the real and imaginary parts of the linear optical conductivities $\sigma^{\pm} = \sigma^{xx}\pm i \sigma^{xy}$ in the circularly polarized basis \cite{Li2013,Li_PRB2013}. From this one can also see that, in contrast to the linear conductivities, it is the imaginary part of $\sigma^{\bar\tau\tau\tau}$ that describes the absorptive processes whereas its real part describes virtual processes.  
%

\subsection{Chemical potential dependence}

\begin{figure*}
		\centering
\includegraphics[scale=1]{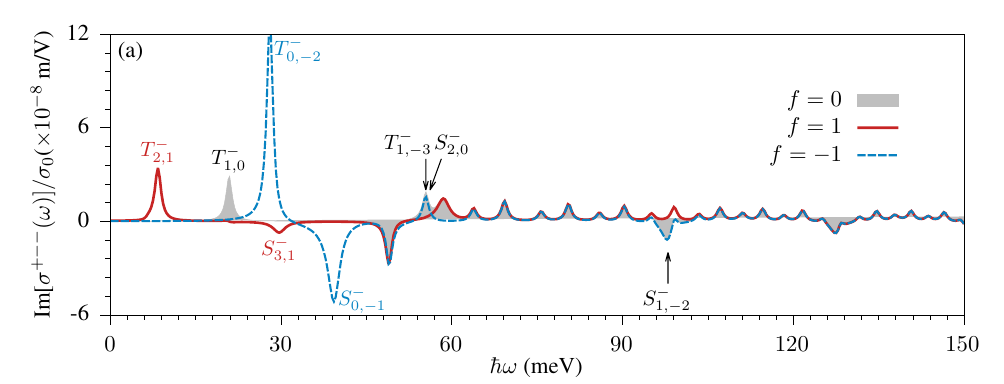}
\includegraphics[scale=1]{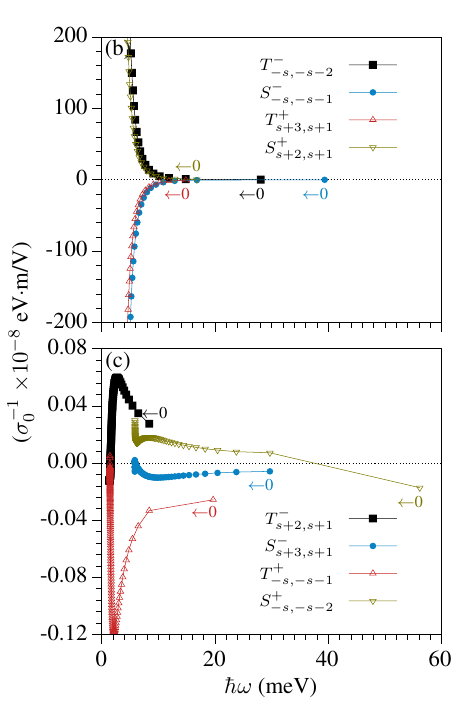}     
\includegraphics[scale=1]{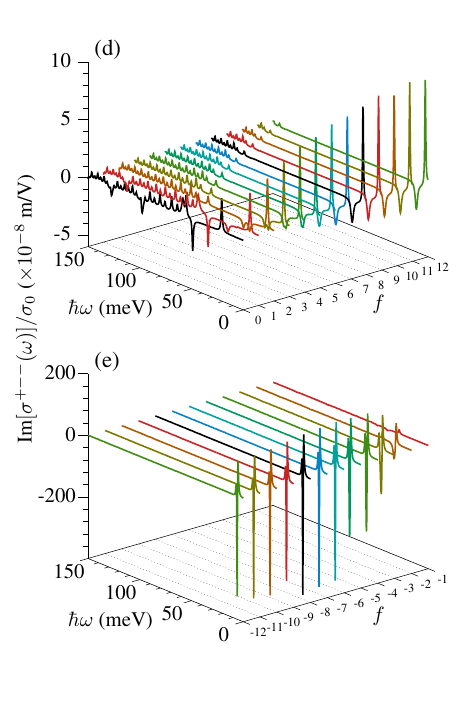}              
\caption{
(a) SHG spectra $\sigma^{+--}$  with the highest filled Landau level $f=-1, 0, 1$. 
Intraband resonant transitions (b) $T^-_{-s,-s-2}$,
$S^-_{-s,-s-1}$, $T^+_{s+3,s+1}$, $S^+_{s+2,s+1}$, (c) 
$T^+_{-s,-s-1}$, $S^+_{-s,-s-2}$, $T^-_{s+2,s+1}$, and  $S^-_{s+3,s+1}$  
for $s \ge 0$ are plotted as functions of the resonant transition energies. The quantities with $s=0$ are marked. 
(d, e) SHG conductivities $\sigma^{+--}$ for different values of $f$.
}
\label{chem}
\end{figure*}

We turn to the effect of chemical potential on the SHG conductivity.
Figure\,\ref{chem}\,(a) illustrates the SHG spectra of Im[$\sigma^{+--}$] for different chemical potentials $\mu=-5,5,50$ meV. Considering the very low temperature and discrete levels, it is more convenient to use the highest occupied Landau levels, $f=-1,0,1$, to represent these chemical potentials. 
Compared to the spectra at $f=0$, which has been shown in Fig.\,\ref{shg}\,(a), the difference of the spectra at different $f$ mainly arises from the Pauli blocking effects: (1) Due to the extra occupied states in conduction bands for $f>0$ or extra unoccupied states in valence bands for $f<0$, some of the interband transitions are blocked and the corresponding peaks disappear. In Fig.\,\ref{chem} (a) these peaks including $T^-_{1,-3}$ and $S^-_{1,-2}$ at $f=1$ and $T^-_{1,0}$ and $S^-_{2,0}$ at $f=-1$ disappear.
(2) There appear new peaks induced by intraband transitions, identified by $T_{2,1}^-$ and $S^-_{3,1}$ at $f=1$, and $T_{0,-2}^-$ and $S^-_{0,-1}$ at $f=-1$. 
To better understand the intraband transitions, figures\,\ref{chem} (b) and (c) show the transition energy dependence of the resonant transition amplitudes $S$ and $T$ at $B=5$ T. 
Unlike the interband transition amplitudes in Figs.\,\ref{shg}\,(c) and (d), here the resonant transition energy decreases with the increasing $s$.
It is interesting to note that for a hole-doped system ($f<0$) the intraband transitions $T^-_{-s,-s-2}$ and $S^-_{-s,-s-1}$ 
have huge amplitudes, which increase with the increasing $s$, while those for electron-doped system ($f>0$) $T^-_{s+2,s+1}$ and $S^-_{s+3,s+1}$ have small values and a complicated $s$ dependence;
 for $S^+$ and $T^+$ the results are opposite. Due to the selection rules, the intraband transition only occurs around the Landau level $f$, leading to only two extra peaks. The huge transition amplitudes are consistent with the Drude contribution, where the Berry connections between the neighbor levels are huge.
Figures\,\ref{chem}\,(d) and (e) show the spectra of $\sigma^{+--}$ as a function of a more extended range of $f$ values.
The behavior of $\sigma^{-++}$ for different $f$ values is similar and is not displayed here.

\subsection{Magnetic field dependence}

\begin{figure*}
		\centering
  \includegraphics[scale=1]{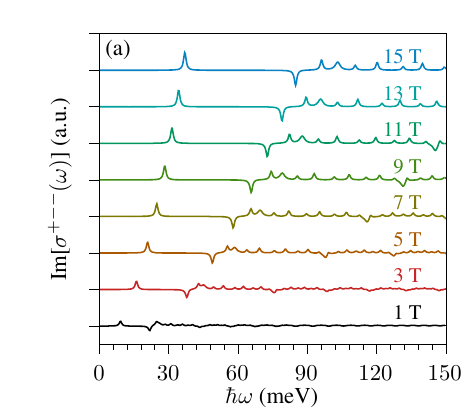}
  \includegraphics[scale=1]{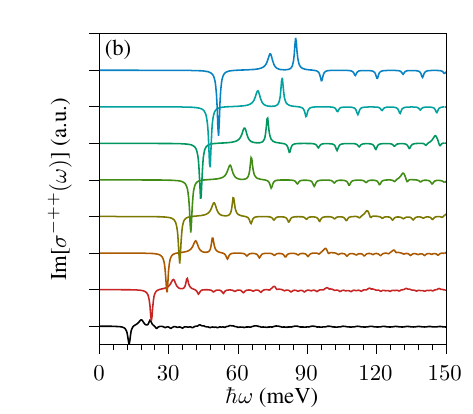}
  \includegraphics[scale=1]{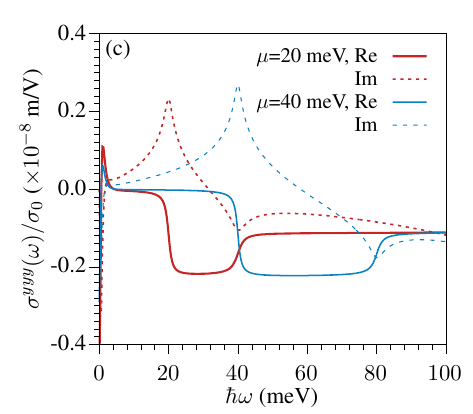}
  \includegraphics[scale=1]{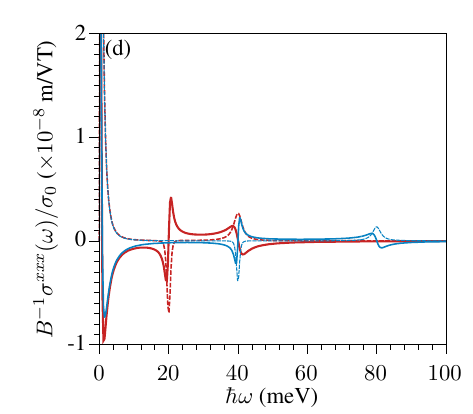}
\caption{Imaginary part of SHG conductivities (a) $\sigma^{(2);+--}$ and (d) $\sigma^{(2);-++}$ with magnetic field intensities. 
All values have been normalized.
Spectra of (c) $\sigma^{(2);yyy}$ and (d) $B^{-1}\sigma^{(2);xxx}$ with chemical potential $\mu = 20, 40$ meV at $B=0.05$ T.
}
\label{intensity}
\end{figure*}

Figures\,\ref{intensity}\,(a) and (b) show the spectra of $\text{Im}[\sigma^{+--}]$ and $\text{Im}[\sigma^{-++}]$ for different magnetic fields varying from $B=$ 1 T to 15 T.
The magnetic field affects the SHG spectra in two aspects: (1) The curves of all SHG spectra show the same structure but the peak locations shift to a higher energy for larger magnetic field, because the eigenenergies of all Landau levels increase with the magnetic field, which results in larger transition energies.
(2) As the magnetic field increases, all peak amplitudes increase, which results in a competition effect between two contributions:
A larger magnetic field leads to larger Landau level energies and smaller Berry connections, further smaller transition amplitudes of $S$ and $T$; however, the larger magnetic field also leads to larger degeneracy of Landau levels, which is shown as the prefactor in Eq.\,(\ref{sigma}). 
The latter dominates and the peak amplitude increases as the magnetic field.

Here it is worth discussing SHG when the magnetic field approaches zero. At exact zero magnetic field $B=0$ the symmetry analysis gives $\sigma^{xxx}(\omega)=0$ and $\sigma^{+--}=-\sigma^{-++}$. 
Thus for small $B$ dependence,  we find that $\sigma^{xxx}(\omega)$ is linearly proportional to $B$, and $\sigma^{yyy}(\omega)$ is approximately independent of $B$. 
Figures\,\ref{intensity}\,(c) and (d) give the spectra of $\sigma^{yyy}(\omega)$ and $B^{-1}\sigma^{xxx}(\omega)$ for two chemical potentials $\mu=20, 40$ meV at magnetic field $B=0.05$ T. We also checked that $\sigma^{yyy}(\omega)$ is an even function of the chemical potential, while $B^{-1}\sigma^{xxx}(\omega)$ is odd.
In this case, 
all spectra show resonant peaks at $\hbar\omega\approx|\mu|$ and $2|\mu|$, associated with the one-photon and two-photon processes, respectively.
In brief, for weak magnetic fields, 
the Landau levels merge to become continuous and the resonant transitions are associated with the chemical potentials, which is consistent with the results in gapped graphene \cite{Cheng2015}.

\section{Conclusion}
\label{conclusion}
In summary, we have investigated the nonlinear magneto-optical conductivity for second harmonic generation of surface states in three-dimensional topological insulators,
where the Landau levels are solved from a two-band hexagonal warping model for the surface states 
under strong external magnetic fields.
Considering the small magnitude of the warping term, a perturbation theory is used to obtain the selection rules for second harmonic generation:
for optical transition from the occupied Landau level $s_2$ to the unoccupied Landau level $s_1$, the selection rules for one-photon transition processes are $|s_1|-|s_2|=\tau$ or $-2\tau$ with $\tau$ being the incident light circularly polarization, while for two-photon resonant transition processes they are $|s_1|-|s_2|=-\tau$ or $2\tau$, encompassing both interband and intraband contributions.
Furthermore, we have also analyzed the amplitude of the resonant transition and revealed the origins of each peak in the spectra.
Based on these results, the behavior of the spectra at different chemical potentials are understood.
We also discussed the magnetic field strength dependence of the SHG.
Specifically, under strong magnetic fields, the peaks become stronger and shift to higher energy with the increase of magnetic field; while in weak magnetic fields, the spectra become continuous, and the $xxx$-direction component is proportional to the magnetic field.
Taking the thickness of the surface states at the order of $d= \hbar v_F/\Delta_{\text{gap}}\approx 9.4$ \AA, which is estimated as the depth of the surface wave functions  penetrating into the bulk  \cite{McIver2012}, the maximal magnitude of SHG susceptibility can reach as high as  $10^7$ pm/V  for light wavelength in the terahertz and infrared frequency regime.
This extremely high value can be tuned by the chemical potential and magnetic field; it exceeds many bulk materials in many magnitudes \cite{Malard2013,Cheng2014,Kumar2013,Guo2005,Seyler2015,Janisch2014}, 
and is comparable to that for twisted bilayer graphene \cite{Liu2020}. 
Our results confirm the unusual nonlinear magneto-optical properties in topological insulators, and propose it as a possible functional nonlinear material in developing novel nonlinear magneto-optical devices.

\begin{acknowledgement}

Work in China was supported by National Natural Science 
Foundation of China Grants No. 12034003,
and No. 62250065 (K.C. and J.L.C.). Work in U.S. was supported by the U.S. Department of Energy, Office of Science, Basic Energy Sciences under Early Career Award No. DE-SC0019326 (W-K.T. and M.Z.). J.L.C. acknowledges the support from the Talent Program of CIOMP.

\end{acknowledgement}


\bibliography{ref}

\end{document}